\documentclass{article}
\usepackage{spconf,amsmath,graphicx}

\usepackage[utf8]{inputenc} 
\usepackage{cite}
\usepackage{color}
\usepackage{amsfonts}
\usepackage{amsthm}
\usepackage{multirow}
\usepackage{pifont}
\usepackage{url}

\usepackage{makecell}

\newcommand*\samethanks[1][\value{footnote}]{\footnotemark[#1]}

\title{Adapting pretrained speech model for Mandarin lyrics transcription and alignment}
\name{
Jun-You Wang$^1$\sthanks{These authors contributed equally to this work.},
Chon-In Leong$^1$\samethanks[1],
Yu-Chen Lin$^1$,
Li Su$^2$,
Jyh-Shing Roger Jang$^1$
}

\address{
$^1$National Taiwan University, Taiwan\\
$^2$Institute of Information Science, Academia Sinica, Taiwan
}

\copyrightnotice{979-8-3503-0689-7/23/\$31.00~\copyright2023 IEEE}
\begin{document}
%
\maketitle
\begin{abstract}
The tasks of automatic lyrics transcription and lyrics alignment have witnessed significant performance improvements in the past few years. However, most of the previous works only focus on English in which large-scale datasets are available. In this paper, we address lyrics transcription and alignment of polyphonic Mandarin pop music in a low-resource setting. To deal with the data scarcity issue, we adapt pretrained Whisper model and fine-tune it on a monophonic Mandarin singing dataset. With the use of data augmentation and source separation model, results show that the proposed method achieves a character error rate of less than 18\% on a Mandarin polyphonic dataset for lyrics transcription, and a mean absolute error of 0.071 seconds for lyrics alignment. Our results demonstrate the potential of adapting a pretrained speech model for lyrics transcription and alignment in low-resource scenarios.
\end{abstract}
\section{Introduction}\label{sec:introduction}
In the field of music information retrieval (MIR), there are two tasks related to the processing of lyrics: automatic lyrics transcription (ALT) and lyrics alignment~\footnote{Also referred to as lyrics-to-audio alignment~\cite{Gupta2019acoustic}.}. Given a polyphonic audio that contains a mixture of the singing voice and possibly other instruments~\cite{Gao2022music, Demirel2021mstre}, ALT aims at recognizing the lyrics sequence that the singer sings. As for lyrics alignment, given a polyphonic audio and the groundtruth lyrics as the input, it focuses on finding the alignment between the lyrics and audio. The desired granularity of alignment varies across previous works, ranging from phoneme-level~\cite{Schulze2021phoneme}, word-level~\cite{Gupta2019acoustic}, to phrase-level~\cite{Fujihara2011lyrics}. 
A system capable of performing these tasks has various applications, such as generating subtitles for karaoke applications. It also provides high-level features for other MIR tasks, such as cover song detection, query by singing/humming, and even singing voice separation~\cite{Schulze2021phoneme}.

In the past few years, with the use of a large-scale dataset~\cite{Meseguer2018dali}, novel neural network architecture~\cite{Gao2022music}, and multitask learning technique~\cite{Huang2022improving, Gao2022automatic}, the performance of both ALT~\cite{Gupta2020automatic, Gupta2018semi, Demirel2020automatic, Demirel2021mstre, Gao2022music, Gao2022automatic, Gao2022genre, Basak2021end} and lyrics alignment~\cite{Stoller2019end, Gupta2019acoustic, Sharma2019automatic, Durand2023contrastive, Huang2022improving} systems have improved significantly.
However, most of the previous works only focused on English, in which a large-scale dataset that contains hundreds of hours of data is available~\cite{Meseguer2018dali}. Due to the amount of data required for training, they are not directly applicable to other low-resource languages. 
A possible solution is to train a model on cross-lingual datasets~\cite{Vaglio2020multilingual, Durand2023contrastive}. By taking advantage of shared characters~\cite{Durand2023contrastive} or phoneme sets~\cite{Vaglio2020multilingual}, these methods benefit from the large-scale English dataset on lyrics alignment in various European languages. However, they still cannot be generalized to languages that do not share many characters or phonemes with English.

Recently, Whisper~\cite{Radford2022robust}, a family of end-to-end speech models for automatic speech recognition (ASR) and speech translation, has demonstrated state-of-the-art performance in multilingual ASR. This opens up another avenue for addressing ALT and lyrics alignment in low-resource languages. In this work, we propose to adapt Whisper medium~\cite{Radford2022robust} for Mandarin lyrics transcription and alignment at character-level, with only a few hours of labeled data. This allows us to assess the feasibility of utilizing Whisper for low-resource ALT and lyrics alignment. It also marks an early attempt at Mandarin ALT and lyrics alignment, which have only been discussed in a few works, including~\cite{Ren2023Transcription} and the MIREX 2018 automatic Lyrics-to-Audio Alignment competition~\footnote{\url{https://www.music-ir.org/mirex/wiki/2018:Automatic_Lyrics-to-Audio_Alignment_Results}}.

In practice, this idea faces two main challenges, including 1) Whisper's architecture may not be suitable for lyrics alignment, and 2) there is a discrepancy between the data used for pretraining and fine-tuning. For the first issue, while Whisper provides the time-aligned transcription mode, its end-to-end architecture that directly predicts discrete timestamps to indicate boundaries may not be robust enough for lyrics alignment. For the second issue, the input data for pretraining Whisper is a speech signal. This is different from the input for ALT and lyrics alignment, which contains a mixture of singing and accompaniments. These issues make it not trivial to adapt Whisper to ALT and lyrics alignment.

To address the first issue, for lyrics alignment, we build an RNN on top of Whisper's encoder. It produces a posteriogram for the input audio. Then, we apply Viterbi forced alignment~\cite{Forney1973viterbi} to determine the alignment between the posteriogram and the lyrics, similar to previous lyrics alignment works~\cite{Stoller2019end, Gupta2019acoustic, Sharma2019automatic, Huang2022improving}. 
As for the second issue, we apply HT Demucs~\cite{Rouard2023hybrid}, a source separation model that extracts vocals from the mixture as a pre-processing step. This reduces the domain gap between the data used for pretraining (speech) and the data used for fine-tuning and testing. Although such a setting is different from most of the recent works which directly operate on the polyphonic audio~\cite{Stoller2019end, Gupta2020automatic, Demirel2021mstre, Durand2023contrastive}, our results show that the pre-processing step does improve the performance on both tasks. 
The results also suggest that such an improvement can only be observed when the source separation model is sufficiently powerful. This further explains why previous studies concluded that using a source separation model does not lead to improved performance~\cite{Gupta2020automatic, Gupta2019acoustic}.

As for evaluation, we manually create the character-level alignment labels for a subset of the MIR-1k dataset~\cite{Hsu2010improvement}. To the best of our knowledge, this is the first polyphonic Mandarin singing dataset with alignment labels.

After trying various settings, our best model achieves a mean absolute error (MAE) of 0.071 seconds on Mandarin character-level lyrics alignment, and a character error rate (CER) of 17.8\% on Mandarin ALT, with the use of only 5.2 hours of labeled data for training. Our results demonstrate the potential of adapting Whisper for ALT and lyrics alignment systems in low-resource scenarios.

\section{Proposed method}\label{sec:method}
\subsection{Problem definition}\label{subsec:problem_definition}

\subsubsection{Mandarin lyrics transcription}
The task of Mandarin ALT can be formulated as follows: given an audio $\mathbf{x}$ which contains the mixture of Mandarin singing and possibly other instruments, the target is to output the lyrics sequence $\mathbf{l}:=(l_1, l_2, \ldots, l_m)$ that the singer sings in the audio, where $m$ is the length of lyrics. In this work, we represent the lyrics using Mandarin characters, the basic components in Mandarin that convey meanings.

\subsubsection{Character-level Mandarin lyrics alignment}
As for Mandarin lyrics alignment, given a mixture audio $\mathbf{x}$ and the groundtruth lyrics sequence $\mathbf{l}:=(l_1, l_2, \ldots, l_m)$ which is the transcription of the audio, the target is to output a sequence $\mathbf{t} \in \mathbb{R}^{m \times 2}$ that represents the onset and offset timing (i.e., the boundary) of each Mandarin character in $\mathbf{l}$.

\subsection{Backbone model}\label{subsec:backbone_model}
We adopt the pretrained Whisper medium model~\cite{Radford2022robust} as the backbone model. Whisper medium is a Transformer model that adopts the encoder-decoder structure. It is the second largest model in the Whisper model family. It was trained on end-to-end multilingual ASR, speech translation, spoken language identification, and voice activity detection. By learning from large-scale datasets in a weakly supervised manner, Whisper medium has demonstrated state-of-the-art performance and high robustness in multilingual ASR. For convenience, we directly refer to the Whisper medium model as Whisper in the remainder of this paper.

\begin{figure}[t]
    \centering
    \includegraphics[width=0.45\textwidth]{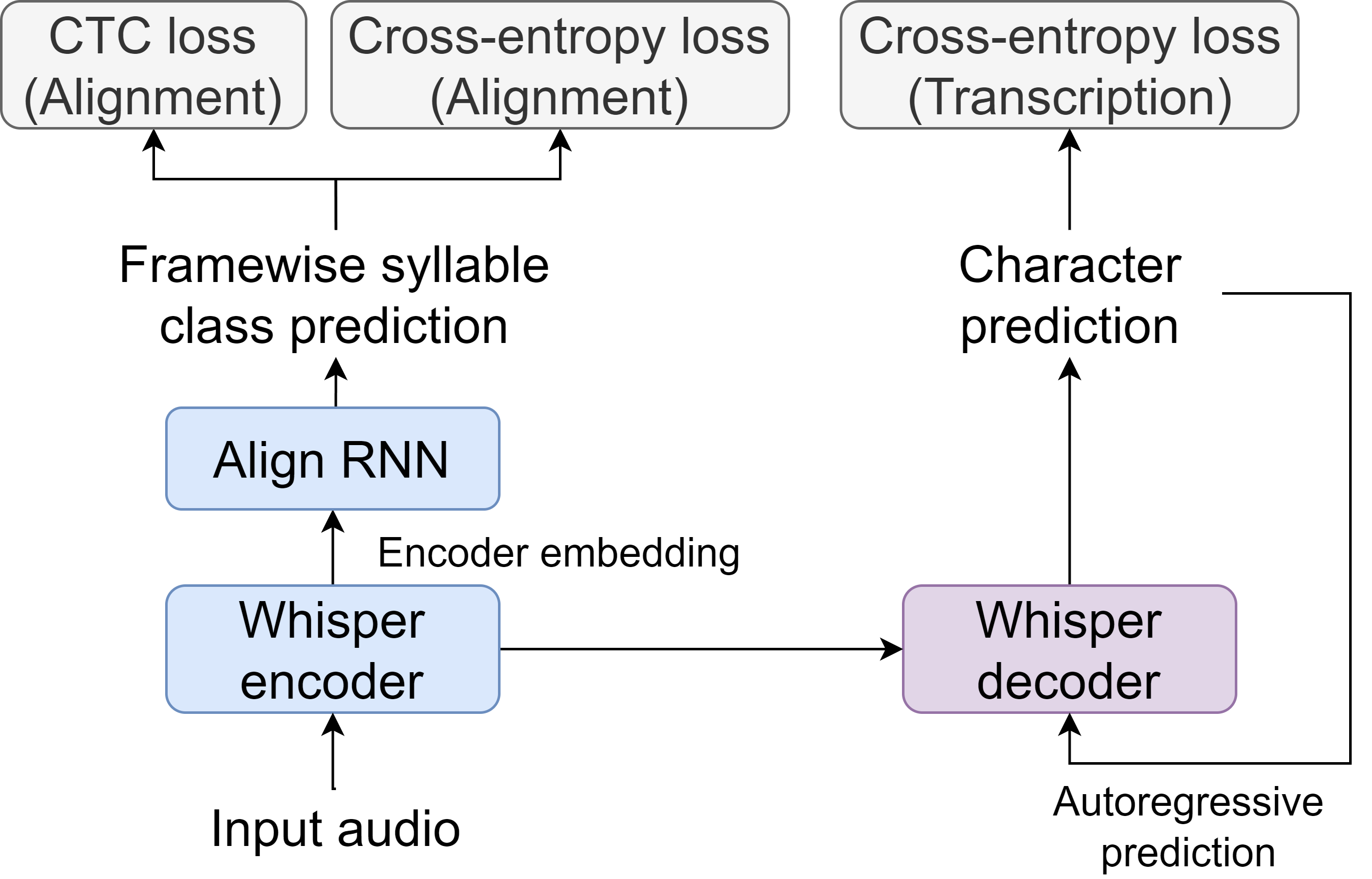}
    \caption{The block diagram of the proposed method for lyrics alignment. During inference, we use the framewise syllable class prediction for lyrics alignment.}
    \label{fig:diagram}
\end{figure}

\subsection{Proposed model}\label{subsec:proposed_model}
We use Whisper~\cite{Radford2022robust} as the backbone model to build the Mandarin ALT and lyrics alignment model.
For Mandarin ALT, as its problem formulation is similar to ASR, we directly adopt the original Whisper model for ALT without making any modifications. During fine-tuning, we train it on Mandarin singing data with the cross-entropy loss similar to the pretraining phase of Whisper.

As for lyrics alignment, a naive way is to use Whisper's time-aligned transcription mode~\cite{Radford2022robust}, which was pretrained to output a discrete timestamp token at the boundary of each voiced segment. It is possible to fine-tune Whisper to output a timestamp at the boundary of each Mandarin character for alignment. However, this approach has a potential issue that the timestamp tokens are discrete, and each of which is treated as an independent class. This may lead to low robustness, which is further emphasized in low-resource scenarios.

Therefore, instead of using Whisper's time-aligned transcription mode, we propose to add an RNN on top of Whisper's encoder, which we denote as align RNN, for lyrics alignment. The align RNN consists of a 2-layer bidirectional GRU and a linear layer, with a Mish activation function~\cite{Misra2020mish} between them. Figure~\ref{fig:diagram} shows an overview of the proposed approach. It predicts a probability distribution for each frame, which is then used for forced alignment to produce the alignment between lyrics and audio. We train align RNN to predict the pronunciation of the Mandarin character that the singer sings at each frame. Compared to directly predicting Mandarin characters, this reduces the number of classes that the model has to distinguish, which could be beneficial in low-resource scenario.

In practice, we use pypinyin~\footnote{https://github.com/mozillazg/python-pinyin} to convert Mandarin characters to pinyin, a romanization system for describing the pronunciation of Mandarin syllables. Each distinct syllable is treated as an individual class, which leads to 401 syllable classes in total. An additional class that represents silence (rest) is also added. During training, we apply the combination of the CTC loss~\cite{Graves2006connectionist} and the cross-entropy loss for optimization. The final alignment loss is the unweighted sum of the two loss functions. During inference, we apply Viterbi forced alignment~\cite{Forney1973viterbi} to align the lyrics and the model's framewise prediction.

For both tasks, we train the model with the unweighted sum of lyrics alignment loss and ALT loss, i.e., applying multitask learning, as shown in Figure~\ref{fig:diagram}.

\subsection{Pre-processing}\label{subsec:pre_processing}
One of the largest challenges in both ALT and lyrics alignment is the possible presence of accompaniments, i.e., the background music, in the input audio. This poses a difficulty in performing ALT and lyrics alignment~\cite{Gupta2018semi, Sharma2019automatic, Gupta2020automatic, Gao2022automatic}.

To address this issue, previous works have employed different approaches. Some works utilize a source separation model as a pre-processing step to remove the accompaniments~\cite{Gupta2018semi, Sharma2019automatic}, while others directly train ALT/lyrics alignment models on the mixture audio~\cite{Stoller2019end, Gupta2020automatic, Demirel2021mstre}. A very recent work further proposed to take both the mixture audio and the extracted vocals as input~\cite{Gao2022automatic}. Previous studies have highlighted both the advantages and disadvantages of using source separation models. On the one hand, since the extracted vocals are closer to clean vocals, it is easier to adapt a model trained on clean vocals or even speech data~\cite{Sharma2019automatic} to it than to the mixture. On the other hand, source separation models may introduce distortions and artifacts, which negatively affect the model performance. Experimental results from previous works have shown that applying source separation models often leads to worse performance~\cite{Gupta2020automatic, Gupta2019acoustic}, which suggests that the disadvantage outweighs the advantage.

However, it should be noted that the experiments in previous works were conducted with source separation models that are less effective than the current state-of-the-art. With the use of a better source separation model, the disadvantage could be largely alleviated. Therefore, we choose to use HT Demucs~\cite{Rouard2023hybrid}, a recently proposed music source separation model which achieved state-of-the-art performance, as a preprocessor to extract the vocals. We then pass the extracted vocals to Whisper for either ALT or lyrics alignment.

As for the approach that uses both the mixture and extracted vocals as the input~\cite{Gao2022automatic}, since Whisper only takes one audio as the input, if we want to adopt this approach, we have to modify Whisper's model architecture. Therefore, we do not choose to use this approach in this work.

\subsection{Dataset}\label{subsec:dataset}
We use the Opencpop dataset~\cite{Wang2022opencpop} and the MIR-1k dataset~\cite{Hsu2010improvement} for experiments. The Opencpop dataset contains 5.2 hours of clean vocals sung by one female professional singer, with lyrics labels and the alignment labels at both the phoneme level and character level~\cite{Wang2022opencpop}. The MIR-1k dataset consists of 2.2 hours of polyphonic recording in which the vocals were sung by 19 amateur singers. The vocals and accompaniments were recorded in separate tracks. The dataset contains the lyrics labels for each clip, but without any character-level alignment label. We first remove all the songs with non-Mandarin lyrics. For the remaining songs, we automatically remove the blanks and convert all the lyrics labels to Simplified Chinese using chinese-converter~\footnote{https://github.com/zachary822/chinese-converter}. For MIR-1k, we also remove the songs which are overlapped with the Opencpop dataset to ensure a fair evaluation.

To evaluate the model performance on lyrics alignment on polyphonic songs, we manually label the character-level alignment for a subset of the MIR-1k dataset, which contains 17 songs and has a total duration of 19 minutes and 33 seconds. This is about one-sixth of the duration of the whole MIR-1k dataset. The annotations are available at \url{https://github.com/navi0105/LyricAlignment}.

For dataset partition, to better evaluate the model performance in an out-of-distribution scenario, which is crucial in evaluating a model's robustness~\cite{Radford2022robust}, we leave the whole MIR-1k dataset for testing. As for the Opencpop dataset, we adopt its official train/test split, which divides the 100-song dataset into a 95-song training set and a 5-song test set.

\subsection{Data augmentation}\label{subsec:data_augment}
Based on the data partition, the training data only contains monophonic audio in Opencpop's training set, which may lead to a domain gap between the training data and test data. Even with the use of HT Demucs~\cite{Rouard2023hybrid}, the extracted vocals still have some differences from the clean vocals. Therefore, a model trained on clean vocals may not perform as well on polyphonic audio. This issue cannot be resolved by simply introducing other polyphonic datasets, since to the best of our knowledge, MIR-1k is the only publicly available polyphonic Mandarin singing dataset with lyrics labels. 

To alleviate this issue, we apply data augmentation by automatically mixing instruments to the vocal tracks in the Opencpop dataset. We use the Musdb dataset~\cite{musdb18} for data augmentation, which contains 150 songs with separated vocal and accompaniment tracks. Considering that HT Demucs was trained with Musdb's training set~\cite{Rouard2023hybrid}, we choose to augment the Opencpop dataset with only Musdb's 50-song test set, so that all the tracks used for augmentation have not been seen by HT Demucs.

In practice, we create three augmented versions of the Opencpop dataset by changing the signal-to-noise ratio (SNR) between the vocal and the augmented accompaniments. The SNR of the three versions are fixed at 0, -5, and -10, respectively. We use all three augmented versions and the original clean vocals for model training.

\section{Experiments and results}\label{sec:experiments_results}
\subsection{Experimental settings}\label{subsec:experimental_settings}
{\bf Model training.} During training, we leave 5\% of Opencpop's training set for validation. For lyrics alignment, we train the model for 2000 steps using the AdamW optimizer with an initial learning rate of $5 \times 10^{-6}$ for the backbone model and $5 \times 10^{-3}$ for align RNN; for ALT, the number of steps is changed to 600, and the initial learning rate for the backbone model is changed to $1 \times 10^{-6}$. The batch size is set to 16. We apply a linear learning rate scheduler with 200 warmup steps. Model validation is performed once every 200 steps.

{\bf Model setting.} We set the hidden dimension of the GRU layers in align RNN to 384, and apply a dropout rate of 0.15 between the GRU layers.

{\bf Evaluation metrics.} For lyrics alignment, we compute mean absolute error (MAE)~\footnote{Also referred to as average absolute error (AAE) in some of the previous work~\cite{Durand2023contrastive}.} in seconds, which is the average absolute error of each predicted character boundary and the corresponding groundtruth label. As for ALT, we adopt character error rate (CER) as the main evaluation metric, which is calculated by dividing the Levenshtein distance between the predicted character sequence and the groundtruth by the length of the groundtruth. In addition, we report the phoneme error rate (PER) as a secondary metric. Since Whisper predicts Mandarin characters in an end-to-end manner, we apply pypinyin~\footnote{https://github.com/mozillazg/python-pinyin} to convert both the predicted and groundtruth characters to phonemes, and then compute the error rate between the converted sequences.

{\bf Test data.} We use Opencpop's test set and the MIR-1k dataset for testing. Since the MIR-1k dataset contains separated vocal and accompaniment tracks, we can either use the mixture of the two tracks as the input or use the vocal tracks as the input. The latter can be viewed as the case where an oracle source separation model is available. To avoid ambiguity, we denote the former one as \emph{MIR-1k mix}, and the latter one as \emph{MIR-1k vocal}.


\begin{table}[t]
\begin{center}
\begin{tabular}{c|cc}
\hline
Model setting & Opencpop test & MIR-1k mix\\
\hline\hline
Whisper decoder & {\bf 0.0193} & 0.8220 \\
\hline
Proposed & 0.0228 & {\bf 0.0709}\\
Proposed (w/o CTC) & 0.0597 & 0.1311\\
Proposed (w/o aug) & 0.0239 & 0.0993\\
Proposed (char) & 0.0449 & 0.1178\\
\hline
\end{tabular}
\caption{The MAE of different model settings on Mandarin lyrics alignment.}\label{tab:align_main}
\end{center}
\end{table}

\subsection{Lyrics alignment results}\label{subsec:alignment_results}
Table~\ref{tab:align_main} shows the lyrics alignment results on \emph{Opencpop test} and \emph{MIR-1k mix}. We compare the performance of 1) \emph{Whisper decoder}, in which we train Whisper's time-aligned transcription mode for lyrics alignment, 2) \emph{Proposed}, which is the proposed method, 
3) \emph{Proposed (w/o CTC)}, which removes the CTC loss, 4) \emph{Proposed (w/o aug)}, which does not apply data augmentation on the training data, and 5) \emph{Proposed (char)}, which trains the model to output the probability distribution of Mandarin characters instead of syllable classes. 

On \emph{MIR-1k mix}, \emph{Proposed} performs the best, achieving an MAE of 0.0709 seconds. This demonstrates the feasibility of adapting Whisper to Mandarin lyrics alignment for polyphonic audio. Considering the scale of the training dataset (about 5 hours), such a result also shows the potential of the proposed method to work in other low-resource languages. Furthermore, the results that \emph{Proposed} outperform other settings on \emph{MIR-1k mix} also confirm the effectiveness of the proposed model settings for lyrics alignment in polyphonic audio, including the design of align RNN, the use of the CTC loss, the use of data augmentation, and the design of outputting syllable classes instead of Mandarin characters.

However, despite underperforming \emph{Proposed} on \emph{MIR-1k mix} by a large margin, \emph{Whisper decoder} slightly outperforms \emph{Proposed} on \emph{Opencpop test}. We suspect that two factors have contributed to such results. First, as discussed in Section~\ref{subsec:proposed_model}, Whisper's time-aligned transcription mode is not robust. Although the model performs well on the in-domain \emph{Opencpop test}, it may not perform as well on out-of-domain data such as \emph{MIR-1k mix}. Second, there is a duration discrepancy between the audios in the Opencpop dataset (mostly under 7 seconds~\cite{Wang2022opencpop}) and the MIR-1k dataset (typically one verse and one chorus). During training, only the timestamp tokens representing time intervals under 7 seconds are well-trained. This may lead to unsatisfactory results on \emph{MIR-1k mix}, which requires timestamp tokens that exceed 7 seconds to represent some of the character boundaries.

To further evaluate the effect of the two factors, we divide the songs in \emph{MIR-1k mix} into multiple segments with an average duration of about 7 seconds and run \emph{Whisper decoder} again. This leads to an MAE of 0.412 seconds. which is better than the original result but is still far behind \emph{Proposed}. This suggests that both factors do contribute to the poor model performance, and also confirms that the proposed method is more robust than Whisper's time-aligned transcription mode when applied to out-of-domain data.\\

\begin{table}[t]
\begin{center}
\begin{tabular}{c|cc}
\hline
Model setting & Opencpop test & MIR-1k mix\\
\hline\hline
Proposed (HT Demucs) & {\bf 0.0228} & {\bf 0.0709}\\
Proposed (Spleeter) & 0.0262 & 0.1231\\
Proposed (mixture) & 0.0249 & 0.0980\\
\hline
\end{tabular}
\caption{The MAE of the proposed model with different preprocessing approaches on Mandarin lyrics alignment.}\label{tab:align_preprocess}
\end{center}
\end{table}

Besides the above settings, we also want to evaluate the effectiveness of the pre-processing step. We compare three methods for pre-processing, including \emph{Proposed (HT Demucs)}, \emph{Proposed (Spleeter)}, and \emph{Proposed (mixture)}. \emph{Proposed (HT Demucs)} denotes the proposed method which we use HT Demucs~\cite{Rouard2023hybrid} for pre-processing, and is exactly the same as \emph{Proposed} in Table~\ref{tab:align_main}. As for \emph{Proposed (Spleeter)}, we also apply a pre-processing step, but use Spleeter~\cite{Hennequin2020spleeter} instead of HT Demucs for source separation. While Spleeter is also a source separation model, its performance falls behind HT Demucs on vocals signal to distortion ratios (vocals SDR) by about 2dB (6.86 vs. 8.93) on Musdb's test set~\cite{Hennequin2020spleeter, Rouard2023hybrid}. Using Spleeter allows us to examine how the performance of source separation models affects the downstream lyrics alignment task. As for \emph{Proposed (mixture)}, we do not apply any pre-processing. Instead, we directly use the original audio for training and testing.

The results are shown in Table~\ref{tab:align_preprocess}. On \emph{Opencpop test}, all the models perform similarly. This fits our expectation as \emph{Opencpop test} contains only clean vocals. As for \emph{MIR-1k mix}, \emph{Proposed (HT Demucs)} performs the best. Interestingly, \emph{Proposed (mixture)} performs the second best. These results suggest that using a source separation model for pre-processing is not always bad or good for lyrics alignment. It depends on how good the source separation model is. When advanced source separation models such as HT Demucs had not been proposed, it is reasonable that previous works concluded that using a source separation model for pre-processing leads to worse performance~\cite{Gupta2020automatic, Gupta2019acoustic}. 

\begin{table}[t]
\begin{center}
\begin{tabular}{c|cc|cc}
\hline
\multirow{2}{*}{\centering Model setting}
& \multicolumn{2}{c|}{Opencpop test} & \multicolumn{2}{c}{MIR-1k mix} \\
\cline{2-5}
& CER & PER & CER & PER\\
\hline\hline
Whisper pretrained & 10.7 & 2.8 & 18.1 & 10.0\\
\hline
Proposed & {\bf 9.3} & {\bf 2.0} & {\bf 17.8} & {\bf 9.6}\\
Proposed (w/o aug) & 9.9 & {\bf 2.0} & 20.6 & 11.3\\
\hline
\end{tabular}
\caption{The ALT performance of different model settings. All the numbers are in percentage points.}\label{tab:transcript_main}
\end{center}
\end{table}

\begin{table}[t]
\begin{center}
\begin{tabular}{c|cc|cc}
\hline
\multirow{2}{*}{\centering Model setting}
& \multicolumn{2}{c|}{Opencpop test} & \multicolumn{2}{c}{MIR-1k mix} \\
\cline{2-5}
& CER & PER & CER & PER\\
\hline\hline
Proposed (HT Demucs) & {\bf 9.3} & {\bf 2.0} & {\bf 17.8} & {\bf 9.6}\\
Proposed (Spleeter) & {\bf 9.3} & 2.3 & 28.0 & 17.5\\
Proposed (mixture) & 10.3 & 2.5 & 25.0 & 15.0\\
\hline
\end{tabular}
\caption{The ALT performance of the proposed model with different preprocessing approaches.}\label{tab:transcript_preprocess}
\end{center}
\end{table}

\subsection{Lyrics transcription results}\label{subsec:transcription_results}
Table~\ref{tab:transcript_main} shows the Mandarin ALT results on \emph{Opencpop test} and \emph{MIR-1k mix}. We compare the model performance of 1) \emph{Whisper pretrained}, which is the pretrained Whisper model without fine-tuning, 2) \emph{Proposed}, which is the proposed method, 
and 3) \emph{Proposed (w/o aug)}, which do not apply data augmentation on the training data. For post-processing, we automatically remove English characters and blanks, and convert all the outputs to Simplified Chinese. We use Whisper's default setting of beam size of 5 for evaluation.

Of all the settings, \emph{Proposed} performs the best on all the datasets. However, the differences are only marginal on \emph{MIR-1k mix}, with only 0.3\% absolute CER and 0.4\% absolute PER over \emph{Whisper pretrained}. This shows the zero-shot ability of Whisper on Mandarin ALT. As for the comparison of \emph{Proposed} and \emph{Proposed (w/o aug)}, similar to the lyrics alignment results, \emph{Proposed} outperforms \emph{Proposed (w/o aug)}, which confirms the effectiveness of applying data augmentation. \\

As for the comparison of different pre-processing settings, we again compare three settings, including \emph{Proposed (HT Demucs)}, \emph{Proposed (Spleeter)}, and \emph{Proposed (mixture)}. The results are shown in Table~\ref{tab:transcript_preprocess}. \emph{Proposed (HT Demucs)} performs the best, while \emph{Proposed (mixture)} performs the second best. This again demonstrates the effectiveness of using HT Demucs as a pre-processor. It also shows that the choice of source separation model largely affects the model performance.

\subsection{Time-aligned lyrics transcription}\label{subsec:time_align_transcription}
In the above experiments, we have trained models for lyrics alignment and ALT. To go one step further, it is possible to combine the two models, which enables the direct generation of time-aligned lyrics from the input audio. We refer to this task as \emph{time-aligned lyrics transcription}, which is analogous to the \emph{time-aligned ASR} in speech recognition. To achieve this task, we first use an ALT model to transcribe lyrics. Then, we feed both the input audio and the predicted lyrics to a lyrics alignment model to generate the character-level alignment.


To evaluate the model performance, we adopt the Percentage of Correctly Aligned Segments (PCAS)~\cite{Fujihara2011lyrics, Schulze2021phoneme} metric. It evaluates the percentages of audio's duration that are labeled correctly. In the context of time-aligned lyrics transcription, the correct label of each segment, i.e., the groundtruth lyrics, is not available and has to be predicted from the input audio.

We evaluate our best model, \emph{Proposed}, on \emph{Opencpop test}, \emph{MIR-1k vocal} and \emph{MIR-1k mix}. For a comprehensive study, we adopt two different criteria to judge if the label is correct, including 1) the Mandarin character should be the same as the groundtruth, and 2) the pronunciation of the Mandarin character should be the same as the groundtruth. We denote the former one as \emph{PCAS-exact}, and the latter one as \emph{PCAS-pronoun}.

\begin{table}[t]
\begin{center}

\begin{tabular}{c|cc}
\hline
Dataset & PCAS-exact & PCAS-pronoun\\
\hline\hline
Opencpop test & 86.58\% & 91.80\%\\
MIR-1k vocal & 85.85\% & 89.54\%\\
MIR-1k mix & 77.26\% & 81.62\%\\
\hline

\end{tabular}
\caption{The time-aligned lyrics alignment results of the proposed model (\emph{Proposed}) on various datasets.}\label{tab:time_aligned_transcription}
\end{center}
\end{table}

Table~\ref{tab:time_aligned_transcription} shows the results. On the polyphonic \emph{MIR-1k mix}, the proposed model achieves a \emph{PCAS-exact} of 77.26\%. In other words, even for polyphonic audio, it can correctly predict the Mandarin character that the singer sings in more than 77\% of the frames, without the need of any groundtruth lyrics. When the input audio only contains vocals, the \emph{PCAS-exact} reaches 86.58\% on \emph{Opencpop test}, and 85.85\% on \emph{MIR-1k vocal}. For all three datasets, the \emph{PCAS-pronoun} results are about 4\% higher than \emph{PCAS-exact}. This represents the cases when the model correctly recognizes the pronunciation of a Mandarin character, but fails to determine the correct character. Overall, these results demonstrate the effectiveness of the proposed model on time-aligned lyrics transcription.

Furthermore, these results also suggest that, despite the advancements in source separation models, even the current state-of-the-art cannot fully eliminate the influence of accompaniments. This leads to the performance differences between \emph{MIR-1k vocal} and \emph{MIR-1k mix}. With the future advancements in source separation models, the performance of ALT, lyrics alignment, and even time-aligned lyrics transcription, can be further improved.

\subsection{Limitations}\label{subsec:limitations}
After the experiments, when we examined the pretrained Whisper's transcription (the \emph{Whisper pretrained} setting), we found that it incorrectly appends some Mandarin characters at the end of its prediction on two of the songs in \emph{MIR-1k mix}. These characters mean ``lyricist and composer: Lee Chung-shan''. Lee Chung-shan is a musician who writes Mandarin pop music. Interestingly, both songs in \emph{MIR-1k mix} were not written by him. That is to say, these characters do not make any sense given the context of the input audio.

A similar phenomenon has been reported and discussed in Whisper's original paper~\cite{Radford2022robust}, in which this is referred to as ``complete hallucination''. It is clear that Whisper learned these characters from its training data, which may contain Mandarin songs that were actually written by Lee Chung-shan. If we think one step further, there is also a possibility that other Mandarin songs, including some of the songs we leave for testing, may have also been used for pretraining Whisper. If this is the case, it would affect the experimental results and make us overrate the model performance.

The main issue is that, we can only know from Whisper's paper~\cite{Radford2022robust} that it constructed a dataset from the Internet for training. We do not know whether each song we use for testing is in the dataset or not. If the dataset is publicly available, we can verify the overlapping between it and our test set, and then exclude the overlapped songs. However, this solution is not viable as we do not have access to Whisper's dataset.
Therefore, to avoid possible misleading, we choose to discuss the possibility of data contamination here, and admit that this may be a limitation in our experimental results. 

In addition to the data contamination issue, we would also like to point out that, while our experiments are conducted in a low-resource setting where only 5.2 hours of Mandarin data are used, during Whisper's pretraining, Mandarin (Chinese) is not a low-resource language. Based on Whisper's paper~\cite{Radford2022robust}, their dataset contains 23,446 hours of Chinese speech data for multilingual ASR. It is unclear whether the results in our work can be replicated in other languages where Whisper did not have access to such a scale of data for pretraining. This requires the construction of polyphonic datasets for various languages to verify. We leave this as future work.

\section{Conclusions}\label{sec:conclusions}
In this paper, we have demonstrated the effectiveness of adapting Whisper on Mandarin ALT and lyrics alignment. With the help of a source separation model, the use of data augmentation, and a better model setting, the proposed Mandarin ALT and lyrics alignment model achieve promising results on polyphonic data, with the access of only 5.2 hours of singing data for fine-tuning. These results suggest that adapting pretrained speech models on singing has the potential of working on low-resource scenarios, where only a few hours of labeled singing data are available. This could be beneficial for the music industry in multiple languages where the lack of data has long hindered the development of ALT and lyrics alignment systems.

\section{Acknowledgements}
We would like to thank Meng-Hua Yu for the efforts in annotating the MIR-1k dataset.

\bibliographystyle{IEEEbib}
\bibliography{ASRU2023}

\end{document}